\documentstyle[11pt]{article}

\def\thefootnote{\fnsymbol{footnote}}
\def\roughly#1{\mathrel{\raise.3ex\hbox{$#1$\kern-.75em%
\lower1ex\hbox{$\sim$}}}}
\def\bm#1{\mbox{\boldmath$#1$\/}}

\def\lsim{\roughly<}
\def\gsim{\roughly>}
\def\be{\begin{eqnarray}}
\def\ee{\end{eqnarray}}

\def\ben{\begin{enumerate}}
\def\een{\end{enumerate}}
\def\beitem{\begin{itemize}}
\def\eitem{\end{itemize}}

\def\thefootnote{\fnsymbol{footnote}}
\newcommand{\beq}{\begin{eqnarray}}
\newcommand{\eeq}{\end{eqnarray}}
\def\la{\langle}
\def\ra{\rangle}
\def\bi{\begin{itemize}}
\def\ei{\end{itemize}}

\def\del{\partial}
\def\L{{\cal L}}

\long\def\beginomit#1\endomit{}
\def\np{{Nucl. Phys.}}
\def\prl{Phys. Rev. Lett.}
\def\pr {Phys. Rev.}

\def\pl{Phys. Lett.}
\def\L{{\cal L}}

\def\chpt{$\chi$PT}

\begin{document}


\begin{titlepage}\begin{center}


\hfill{nucl-th/9507028}

\hfill{June 1995}
\vskip 0.4in
{\Large\bf From Chiral Mean Field to Walecka Mean Field}
\vskip 0.1cm
{\Large\bf And Kaon Condensation}
\vskip 1.2in
{\large  G.E. Brown$^a$\footnote{Supported by the Department of Energy
under Grant No. DE-FG02-88ER 40388} and Mannque Rho$^{b,c}$}\\
\vskip 0.1in
{\large a) \it Department of Physics, State University of New York,} \\
{\large \it Stony Brook, N.Y. 11794, USA. }\\
{\large b) \it Service de Physique Th\'{e}orique, CEA  Saclay}\\
{\large\it 91191 Gif-sur-Yvette Cedex, France}\\
{\large c) \it Institute for Nuclear Theory, University of Washington}\\
{\large \it Seattle, WA 98195, U.S.A.}
\vskip .6in
\vskip .6in

{\bf ABSTRACT}\\ \vskip 0.1in
\begin{quotation}
A unified treatment of normal nuclear matter described by Walecka's mean
field theory
and kaon condensed matter described by chiral perturbation theory
is proposed in terms of mean fields of an effective chiral
Lagrangian. The BR scaling is found to play a key role in making the link
between the ground state properties of nuclear matter and the fluctuation
into the strangeness-flavor direction. A simple prediction for kaon
condensation is presented.
\end{quotation}
\end{center}\end{titlepage}


\renewcommand{\thefootnote}{\#\arabic{footnote}}

\indent
Kaon-nuclear physics is probably one of the most exciting new directions
in nuclear physics. The issues involved are strangeness productions
in heavy-ion collisions and kaon condensation
in compact star matter \cite{br95}.
In discussing kaon-nuclear (or in general pseudo-Goldstone boson) interactions
appropriate for kaon condensation, chiral Lagrangians are found to be
useful in describing
the fluctuation in the strangeness direction. While chiral perturbation theory
(\chpt) is believed to be the most efficient way to implement chiral symmetry
of QCD at low energy, there is a subtlety in applying it to nuclear or
compact star matter which has not yet been satisfactorily addressed in the
literature. This has to do with the consistency in the description of
the ground state of the dense matter and of the fluctuation around that
background defined by the ground state. In treatments to date, there is
no consistency between the two sectors. For instance, the ground state --
nuclear matter -- is successfully
described by Walecka mean-field theory \cite{walecka} with the parameters
of an apparently non-chiral Lagrangian determined from nuclear matter
properties while the kaon condensation phenomenon,
the physics of which is lodged in an effective action (or potential),
is described by low-order chiral perturbation theory with the parameters of a
chiral Lagrangian determined mostly by free-space data. Communication
between the ground-state sector and the fluctuation sector has been glaringly
missing.
What would be needed is a formalism that describes both sectors from a given
chiral Lagrangian with common parameters operative in both sectors
simultaneously. An attractive possibility is that a lump of nuclear matter
arises as a nontopological soliton in the form of ``chiral liquid" as
suggested by Lynn \cite{lynn} around which fluctuations in various flavor
directions could be described. Such a scheme would incorporate chiral symmetry
in a self-consistent way in both sectors \cite{br95,mrpr}. Unfortunately
such a strategy has not been yet formulated into a workable scheme.

In this Letter, we supply the missing link at mean field level with the
help of the BR scaling introduced by us sometime ago \cite{br91}.
\subsection*{Linking chiral and Walecka mean fields}
\indent\indent
We begin by restating the result of \cite{br91} in a form suitable
for making contact with Walecka mean-field theory for nuclear matter.
The starting point of Ref.\cite{br91} is a chiral Lagrangian for large
$N_c$ (where $N_c$ is the number of colors) implemented with the scale
anomaly of QCD, written in terms of pseudo-Goldstone fields $U$, the
scalar field $\chi$ of the scale anomaly and hidden-gauge fields $V_\mu$.
With such a Lagrangian, baryons arise as topological solitons (skyrmions
\cite{skyrmions}). Suppose we embed this chiral Lagrangian into a medium
characterized by density $\rho$ and call it
$\L_\chi$. We assume that $\L_\chi$ incorporates physics above the
chiral scale $\Lambda_\chi\sim 1\ {\mbox{GeV}}$ into
the parameters and counterterms
of the Lagrangian and in a potential $V (U,\chi,V_\mu)$ of the Coleman-Weinberg
type. The basic premise of \cite{br91} is that at a given background density
$\rho$, the chiral Lagrangian takes formally the same form dictated
by QCD symmetries as in matter-free
space with however
the basic parameters of the Lagrangian, $f_\pi^\star$ (pion decay
constant), $g$ (hidden local symmetry constant) and the ``vacuum condensate"
$\la \chi \ra^\star$ etc.  determined at a given density by a
Coleman-Weinberg-type mechanism. This leads at mean field level to the BR
scaling
\be
\frac{m_N^\star}{m_N}\approx \frac{m_V^\star}{m_V}\approx \frac{m_\sigma^\star}
{m_\sigma}\approx\cdots\approx \frac{f_\pi^\star}{f_\pi}\label{brscaling}
\ee
where the subscripts $N$, $V$ and $\sigma$ stand, respectively, for the
nucleon,
the vector mesons ($\rho$, $\omega$) and the scalar meson (to be defined
below). In doing this in a medium, we of course lose manifest Lorentz
invariance
but this is not a serious matter since we can work with Lorentz covariance
as in heavy-quark QCD by using the four-velocity $v_\mu$.
As explained in \cite{br95,ab}, the scalar field $\chi$ that enters
into the medium-dependent constants is the {\it low-energy quarkish component
with the gluonic component having been integrated out}. A similar idea of
separating the trace anomaly into a quarkish component and a gluonic component
has recently been proposed by Furnstahl, Tang and Serot \cite{tang}.
It should
be noted that $\L_\chi$ with (\ref{brscaling}) is to be used in tree order
for describing fluctuations around the background state.
In arriving at this result, we have
used the notion of the baryons as chiral solitons and the consequences,
i.e. low-energy theorems of hidden gauge symmetry \cite{bando}
such as the KSRF relation.

But how is the ground state to be described in this framework? To answer this
question, we approach the same effective theory from a different angle.
For this, we shall exploit a
recent development \cite{mattis,manohar95} that
indicates that the above description using a bosonic Lagrangian
is equivalent at long wavelength
to an effective Lagrangian that contains baryons explicitly
as matter fields. The key idea can be illustrated with
the scale-invariant part of the chiral Lagrangian written in
terms of the Dirac nucleon $\psi$, the pion (or kaon) $U\equiv \xi^2$
and the scalar $\chi$ field:
\be
{\cal L}_0 &=&
{\bar N} \left[ i\gamma^\mu D_\mu
+ i g_A \gamma^\mu\gamma_5 \Delta_\mu
- m_N - g_\chi \chi
\right] N
\nonumber \\
&+& \frac12 (\del_\mu \chi)^2 - \frac12 m_\chi^2 \chi^2 + \cdots
\label{L0}\ee
where the covariant derivatives $\Delta_\mu$ and $D_\mu$ are given by
$\Delta_\mu =
\frac12 \left\{ \xi^\dagger , \del_\mu \xi\right\}
- \frac{i}{2} \xi^\dagger ({\cal V}_\mu+{\cal A}_\mu) \xi
+ \frac{i}{2} \xi ({\cal V}_\mu - {\cal A}_\mu)\xi^\dagger$,
$D_\mu = \del_\mu +\Gamma_\mu$
with $\Gamma_\mu = \frac{1}{2} \left[\xi^\dagger,\, \del_\mu \xi\right]
-\frac{i}{2}\xi^\dagger ({\cal V}_\mu+{\cal A}_\mu) \xi - \frac{i}{2}
\xi ({\cal V}_\mu - {\cal A}_\mu)\xi^\dagger$,
and ${\cal V}_\mu$ and ${\cal A}_\mu$ represent, respectively,
the external vector and axial vector fields introduced specifically
here for reasons that will be apparent shortly.
The ellipsis in (\ref{L0}) denotes terms containing only meson fields
and other meson-baryon couplings that do not concern us for the moment,
As it stands, this Lagrangian can be interpreted as describing
fluctuations around the ``background" defined by the medium with
the scalar field
shifted around that background. We shall show later that the scalar excitation
carries the scaled mass. For the moment, we do not worry about this matter
and take the scalar excitation to be sufficiently massive so
that we can integrate it out. We do this in heavy-baryon formalism
useful for \chpt \ applied to many-nucleon systems \cite{pmr2}:
\be
\delta_\chi {\cal L} &=&
\frac{g_\chi^2}{2 m_\chi^2} \left({\bar B} B\right)^2
+ \frac{g_\chi^2}{2 m_\chi^2\,m_N^2} \bar{B} B
\, {\bar B} \Gamma_{\frac{1}{m_N}} B\nonumber\\
&-& \frac{g_\chi^2}{2 m_\chi^4} \left({\bar B} B\right)
\del^2  \left({\bar B} B\right) +\cdots\label{deltaL}
\ee
where $B$ is the heavy-baryon field and
\be
\Gamma_{\frac{1}{m_N}} =
S^{\mu\nu} \Gamma_{\mu\nu}
+\ 2g_A \left\{v\cdot \Delta, S\cdot D\right\}
- D^2 + (v\cdot D)^2   + g_A^2 (v\cdot \Delta)^2
+\cdots
\label{m1}\ee
where $\Gamma_{\mu\nu}=\frac{\tau_a}{2}\Gamma^a_{\mu\nu}=
\del_\mu \Gamma_\nu- \del_\nu \Gamma_\mu +[\Gamma_\mu,\Gamma_\nu]$
and $S^{\mu\nu}=[S^\mu,S^\nu]$ with $S^\mu$ the spin polarization
four-vector.

Equation (\ref{deltaL}) tells us two things. On the one hand, the first
four-baryon interaction term is one of the four-Fermi interactions expected
in chiral Lagrangians \cite{wein}. In mean field, it is equivalent to
the scalar mean-field potential in Walecka theory \cite{gelmini}.
{\it We should emphasize that the scalar field that enters here is a
chiral singlet, not the fourth component of the scalar quartet of the
linear sigma model.} On the other hand, the second term of (\ref{deltaL})
gives a medium correction to the single-particle axial-charge operator $A^0$
and magnetic moment operator $\bm {\mu}$
in nuclei \cite{pmr2} that shifts the nucleon mass $m_N$ to
$m_N^\star$
(here and in what follows, we will affix $\star$ for quantities defined
in the ``background" field which in our case is characterized by density
$\rho$)
\be
\frac{m_N^\star}{m_N}=(1+\eta\frac{\rho}{\rho_0})^{-1}\label{nucleon}
\ee
giving a correction of the form
\be
\frac{\delta_\chi A^0}{A^0} = \frac{\delta_\chi {\bm \mu}}{{\bm \mu}}
= \eta \frac{\rho}{\rho_0}\label{corr}
\ee
with $\eta=g_\chi^2\rho_0/(m_N m_\chi^2)$.
The two phenomena cited above represent the same physics through
Ward identities. We should note however that while the correction
(\ref{corr}) for the axial charge operator turns out to
remain significant as shown in \cite{KR},
the corresponding correction for the magnetic moment is nearly completely
canceled by the ``back-flow" correction imposed by Galilean invariance, i.e.
the effect of the Fermi-liquid parameter $F_1$.

The other mean field that figures importantly in Walecka theory is the
$\omega$ meson field, $\omega_\mu$, which is {\it also}
a chiral singlet. In chiral
Lagrangian, this couples to the baryon as
\be
\L_{\omega NN}=g_\omega\bar{\psi}\gamma_\mu\psi \omega^\mu \simeq
g_\omega\bar{B}v_\mu B\omega^\mu.\label{deltaLo}
\ee
Again integrating out the $\omega$ field, one gets the additional
piece in the Lagrangian
\be
\delta_{\omega}\L=-\frac{g_\omega^2}{2m_\omega^2} (\bar{B}v_\mu B)^2
+\cdots\label{omegaL}
\ee
Note that (\ref{omegaL})
is yet another chirally symmetric four-Fermi interaction allowed
in chiral Lagrangians. Thus Walecka's repulsive vector mean field can be
identified by a four-Fermi chiral Lagrangian involving $v_\mu$.

The $\rho$ and $a_1$ mesons could be introduced similarly
using hidden local symmetry and generate isospin-dependent mean fields
which however do not figure in symmetric nuclear matter.

We should stress that for mean-field calculations, there is no need to
integrate out the heavy fields: One could work equally well with
the mean fields of the mesons directly as in Walecka's formulation.

The next question we must address is how the coupling constants and
meson masses are defined in the background field. Let us first look at
the vector mean field. Invoking the KSRF relation
${m_V^\star}^2=2{f^\star_\pi}^2 g^2$
where $g$ is the hidden gauge coupling and $V=\rho, \omega$ and
$g_\omega/3=g_\rho=g/2$,
we can write the mean-field potential given by (\ref{omegaL}) as
\be
V_N=\frac{9}{8{f^\star_\pi}^2}\rho.\label{V}
\ee
Analysis in Walecka theory indicates that at nuclear matter density
$\rho=\rho_0$, the vector potential (\ref{V}) has the value
\be
V_N (\rho_0)\approx (0.6)^{-1}\frac{9\rho_0}{8f_\pi^2}\label{vn}
\ee
with $f_\pi=93$ MeV. Thus we see that
\be
\frac{{f^\star_\pi}^2}{f_\pi^2}\approx 0.6.\label{scaling}
\ee
One may understand this scaling from the Gell-Mann-Oakes-Renner (GMOR)
relations.  Assuming that the GMOR relation holds in medium, we
have\footnote{In some approaches, e.g., chiral perturbation theory in
medium, a distinction is made between the ``time component"
and ``space component" of $f_\pi$. Here there is no such distinction;
the fact that we are dealing, in medium, with Lorenz covariance rather than
invariance would lead to an additional (noninvariant) term to (\ref{GMOR}).}
\be
\frac{{f^\star_\pi}^2}{f_\pi^2}\approx \frac{{m^\star_\pi}^2}{m_\pi^2}
\frac{\langle \bar{q}q\rangle^\star}{\langle \bar{q}q\rangle}.\label{GMOR}
\ee
{}From Feynman-Hellman theorem, we have \cite{cohen}
\be
\frac{\langle \bar{q}q\rangle^\star}{\langle \bar{q}q\rangle}
\approx 1-\frac{\Sigma_{\pi N}\rho}{f_\pi^2 m_\pi^2} +\cdots\label{GMOR2}
\ee
where $\Sigma_{\pi N}=45\ {\mbox{MeV}}$ is the $\pi N$ sigma term.
The pion mass is known to change slightly as density increases,
so putting into Eq.(\ref{GMOR}) the empirical value obtained from
$\pi$-mesic atoms \cite{ericson}, $m_\pi^\star (\rho_0)/m_\pi
\approx 1.05$, one obtains (\ref{scaling}).
Since the coupling $g_\omega$ related to the hidden gauge coupling
$g$ does not scale at mean field \cite{br95,br91},
this result immediately implies that
at $\rho\approx \rho_0$
\be
\frac{{m^\star_\omega}^2}{m_\omega^2}=\frac{{m^\star_V}^2}{m_V^2}
= \frac{{f^\star_\pi}^2}{f_\pi^2}\approx 0.6.
\ee

Let us now turn to the scalar meson sector. The role of a low-mass
scalar meson figuring importantly in nuclear phenomenology is much less
clear in the context of chiral perturbation theory. The scalar in the
linear sigma model, namely the fourth component of $O(4)$ group, while
relevant at the critical point of chiral phase transition, is not
visible in low-density regime. The scalar associated with the trace anomaly
is much too massive, $\sim 2$ GeV, to be relevant without substantial
modification. Our proposal is to take
the relevant (chiral singlet)
scalar field to interpolate $2\pi$, $4\pi$ etc. excitations with an effective
low mass $m^\star_\sigma$ (where we have renamed the light (quarkish)
$\chi$ field by $\sigma$ following the convention). We have support for
this idea from the phenomenological meson-exchange model of Durso, Kim and
Wambach \cite{durso}. As reviewed in \cite{br95}, the correlation of the
two pions (and other multi-pion correlations) responsible for the
effective scalar exchange is chiefly accomplished by
crossed-channel $\rho$-exchange.
Inclusion of the density dependence of the $\rho$-meson mass leads to
important modifications in the attraction \cite{durso}. {}From the $\pi\pi
\rightarrow N\overline{N}$ helicity amplitudes $f_+^{J=0} (t)$ supplied
by the authors of \cite{durso}, we can track the downward movement, in mass,
of the (distributed) scalar strength as the density increases. These
$f_+^{J=0} (t)$ helicity amplitudes show a relatively sharp resonance at
$m_\sigma^\star\simeq 500 - 600$ MeV for $\rho\approx \rho_0$. It seems
clear that the strength in the effective scalar degree of freedom from
correlated pion exchange originates in the $\gsim 900$ MeV range at zero
density, and that the exchanged scalar, which is fictitious at $\rho=0$,
materializes as real resonance by $\rho\sim \rho_0$.

What is determined in the phenomenology of Walecka's
mean-field theory is the ratio $(g_\sigma^2/m_\sigma^2)$
in (\ref{deltaL}):
\be
S_N=-\frac{{g^\star_\sigma}^2}{{m^\star_\sigma}^2}\rho_s\label{sn}
\ee
where $\rho_s$ is the scalar density.
Numerically it is determined at $\rho=\rho_0$ to be \cite{brockman}
\be
{g^\star_\sigma}^2/{m^\star_\sigma}^2\approx 5/m_\pi^2.\label{sigmaphen}
\ee
To proceed, let us assume as in \cite{br95} that the nonlinear
chiral Lagrangian ``linearizes" at nuclear matter density. Briefly
what this means is as follows. At zero density and low energies, the strong
interaction is governed by a nonlinear chiral Lagrangian implemented
with the trace anomaly of QCD. As density increases or at higher energy,
the scalar of the trace anomaly tends to be ``dilatonic" and moves towards
the Goldstone pions to form a quartet, thereby approaching an $O(4)$ symmetry.
Even if the mass of the scalar is not degenerate with that of the pions,
the symmetry is ``mended" in the sense described by Beane and van Kolck
\cite{beane}, with the physics of the quartet described effectively by a linear
sigma model. Note, however, that in a strict sense, the scalar will remain
chiral singlet all the way to the chiral restoration point at which
the scalar will merge into the multiplet of $O(4)$. (See \cite{chphase}
on this matter.) Assuming this, we use
the fact that $g_A^\star\approx 1$ at $\rho\approx \rho_0$
and the in-medium Goldberger-Treiman relation to arrive at
\be
g^\star_\sigma\approx 10.
\ee
Therefore (\ref{sigmaphen}) gives
\be
m^\star_\sigma\approx 600 \ {\mbox{MeV}}
\ee
essentially the mass used in mean field calculations.
This implies that
\be
m_\sigma^\star/m_\sigma\approx f_\pi^\star/f_\pi.
\ee

It is not obvious how the nucleon scaling (\ref{nucleon}) is related
to the scaling of the pion decay constant. In the skyrmion description,
one finds \cite{mr88xxx}
$\frac{m^\star_N}{m_N}\approx \sqrt{\frac{g_A^\star}{g_A}}
\frac{f_\pi^\star}{f_\pi}$. Now up to $\rho\approx \rho_0$, in-medium
loop corrections lead to $g_A^\star\approx 1$, and beyond $\rho_0$, $g_A$
remains unscaled. Thus at mean field level, it seems reasonable to take
$\frac{m^\star_N}{m_N}\approx \frac{f_\pi^\star}{f_\pi}$, modulo an overall
constant.

\subsection*{Fluctuations in the kaon sector}
\indent\indent
Consider the S-wave $KN$ interaction relevant to $K$-nuclear interactions
in symmetric nuclear matter. We shall focus on the Weinberg-Tomozawa
term and the sigma term
\be
\L_{KN}=\frac{-6i}{8f^2}(\overline{B}\gamma_0 B)\overline{K}\del_t K +
\frac{\Sigma_{KN}}{f^2}(\overline{B}B)\overline{K}K\equiv {\cal L}_\omega
+{\cal L}_\sigma\label{kaonL}
\ee
where $K^T=(K^+ K^0)$. The constant $f$ in (\ref{kaonL}) can be identified
in free space with the pion decay constant $f_\pi$. In medium, however,
it can be modified as we shall see shortly.
In chiral perturbation expansion, the first term corresponds to
${\cal O} (Q)$ and the second term to ${\cal O} (Q^2)$. There is one more
${\cal O} (Q^2)$ term proportional $\del_t^2$ which we will discuss later.

One can associate the first term of (\ref{kaonL})
arising from integrating out the $\omega$
meson as in the baryon sector. The resulting $K^- N$ vector potential in medium
can then be deduced in the same way as for $V_N$:
\be
V_{K^\pm}=\pm\frac{3}{8{f^\star_\pi}^2}\rho.
\ee
Thus in medium, we may set $f\approx f_\pi^\star$ and obtain
\be
V_{K^\pm}=\pm\frac 13 V_N.\label{omegascale}
\ee
One way of understanding this result is that the constituent quark
description with the $\omega$ meson coupling to the kaon
as a {\it matter field} (rather than a Goldstone) becomes applicable:
the $\omega$ coupling to the kaon which has one nonstrange
quark is 1/3 of the $\omega$ coupling to the nucleon which has three
nonstrange quarks.

Now what about the second term of (\ref{kaonL})?

As in the baryon sector, one may couple the scalar $\chi$ field to the
kaon field. But with the kaon considered as a pseudo-Goldstone field, the
only way to couple them without using derivatives is through the symmetry
breaking term, thus the appearance of the sigma term. On the other hand,
if the kaon is treated as a {\it matter field}
as suggested by its coupling to the
$\omega$-meson field -- a reasonable assumption in medium in view of the fact
that {\it all} non-strange hadron masses other than Goldstone bosons
drop -- then we can use the coupling
\be
\L_\sigma =  \frac 13 2 m_K g^\star_\sigma \overline{K}K\chi
\ee
where the factor 1/3 accounts for one non-strange quark in the kaon as compared
with three in the nucleon.
When the $\chi$ field is integrated out as above, we will get,
analogously to the nucleon case,
\be
\L_\sigma = 2m_K \frac 13
\frac{{g^\star_\sigma}^2}{{m^\star_\sigma}^2}\overline{B}B \overline{K}K.
\ee
Comparing with the second term of (\ref{kaonL}), we find
\be
\frac{\Sigma_{KN}}{f^2}\approx 2\frac{m_K}{3}
\frac{{g^\star_\sigma}^2}{{m^\star_\sigma}^2}.\label{relation}
\ee
This somewhat strange relation may be understood
by recalling the dual role that the kaon plays in the structure of hyperons:
On the one hand, the kaon is light enough to be considered as a
pseudo-Goldstone
boson so the skyrmion picture with $SU(3)$ collective coordinates
applies; on the other hand, it
can be considered as heavy enough
so that it can be treated as a matter field (rather
than a Goldstone field) as in the Callan-Klebanov model \cite{mrpr}.
Reference \cite{MOPR} describes how these two descriptions can be
joined smoothly in the case of the kaon and heavier mesons. It is
shown there that as the mass of a pseudoscalar boson $\Phi$
(such as the $D$ meson) increases above the
chiral scale $\sim 1$ GeV, the heavy-meson field can be taken to transform
like a massive matter field, eventually exhibiting
heavy-quark symmetry. We believe it is this
dual character of the kaon that is reflected in (\ref{relation})
in the background where the masses
of other hadrons scale down as proposed in \cite{br95}: the
Goldstone boson nature of the kaon gives the left-hand side of
Eq.(\ref{relation})
while the massive matter field character of the kaon in medium supplies the
right-hand side. The situation resembles the overlapping in the strange-quark
hadron sector of the ``top-down"
approach starting from heavy-quark symmetry
and the ``bottom-up" approach starting from  chiral symmetry as
discussed in \cite{MOPR}. In the present case, the overlapping
region seems to be located between $\sim \rho_0$ and $\sim 2\rho_0$ in density.

Given the $KN$ sigma term, we could determine independently
what $f$ is. However
since the strange-quark mass is not well-known, we cannot
determine the sigma term accurately even if the $\bar{s}s$ content of
the nucleon could be measured on lattice. Suppose we take it from lattice
gauge calculations \cite{fukugita} (or from
chiral perturbation estimate coming from $KN$ scattering \cite{LBMR}),
$\Sigma_{KN}\approx 3.2 m_\pi$. Then the left- and right-hand sides of
(\ref{relation}) are equal if we set
\be
f\approx f_\pi^\star.
\ee
This is consistent with the scaling for the $\omega$-meson exchange leading
to the relation (\ref{omegascale}).
It follows from (\ref{kaonL}) with $f$ replaced by $f^\star_\pi$
that the scalar kaon-nuclear potential is
\be
S_{K^\pm}=\frac 13 S_N.
\ee
\subsection*{Kaon-nuclear potential}
\indent\indent
Given Walecka mean fields for nucleons, we can now calculate the corresponding
mean-field potential for $K^-$-nuclear interactions in symmetric nuclear
matter.
{}From the results obtained above, we have
\be
S_{K^-} +V_{K^-}\approx \frac 13 (S_N-V_N).
\ee
Phenomenology in Walecka mean-field theory gives
$(S_N-V_N)\lsim -600\ {\mbox{MeV}}$ for $\rho=\rho_0$ \cite{walecka}.
This leads to
the prediction that at nuclear matter density
\be
S_{K^-}+V_{K^-}\lsim -200\ {\mbox{MeV}}.
\ee
This seems to be consistent with the result of the analysis in K-mesic atoms
made by  Friedman, Gal and Batty
\cite{friedman} who find attraction at $\rho\approx 0.97\rho_0$ of
\be
S_{K^-}+V_{K^-}=-200\pm 20\ {\mbox{MeV}}.
\ee
\subsection*{Kaon condensation}
\indent\indent
In applying the above mean-field argument to kaon condensation in
compact-star matter to chiral $O(Q^2)$, we need to make two additions
to the Lagrangian (\ref{kaonL}). The first is that the compact-star matter
relevant to the problem consists of roughly 85\% neutrons and 15\%
protons, so we need to take into account the $\rho$-meson exchange in addition
to the $\omega$ exchange. The $\rho$ exchange between $K^-$ and neutrons
(protons) gives repulsion (attraction), equal in magnitude to 1/3 of
the attraction from the $\omega$-exchange. The second correction has to do
with an $O(Q^2)$ term in the Lagrangian proportional to $\omega_K^2$
(where $\omega_K$ is the kaon frequency) associated with ``1/m" corrections
in chiral expansion. The effect of this term is to cut down the
scalar exchange given by the second term of (\ref{kaonL}) by the
factor $F\simeq (1-0.37\frac{\omega_K^2}{m_K^2})$.

We are now in position to estimate the
critical density for kaon condensation using
the resulting mean-field theory. Let us start with Walecka mean fields
for nucleons in symmetric nuclear matter. We take for illustration
the following values from (\ref{vn}) and (\ref{sn}):
\be
V_N (\rho=\rho_0)\approx 275\ {\mbox{MeV}}, \ \ \
S_N (\rho=\rho_0)\approx -350\ \mbox{MeV}.
\ee
These are, in fact, the values calculated from the Bonn two-body potential
in a relativistic Brueckner-Hartree-Fock approach \cite{brockman} and, thus,
have some grounding in a microscopic theory. As Brockmann and Machleidt
point out, these are more or less central values for those used in Walecka
mean-field calculations.
Now at $\rho\approx \rho_0$, $F\simeq (1-0.37\frac{\omega_K^2}{m_K^2})\approx
0.86$ and so
\be
S_{K^-} (\rho=\rho_0)\approx -\frac 13\times 350\times 0.86 \ \mbox{MeV}\approx
-100 \ \mbox{MeV}.\label{recommend}
\ee
This is valid for $K^-$ independently of the isospin content of the matter.
For symmetric nuclear matter, our scaling gives
\be
V_{K^-} (\rho=\rho_0)\approx -\frac 13\times 275 \
\mbox{MeV} \approx -92  \ \mbox{MeV}.
\ee
Thus for $K^-$ in nuclear matter, we get $(S_{K^-}+V_{K^-})\approx -192 \
\mbox{MeV}$
which is consistent with the K-mesic atom data. For a matter of 85\% neutrons
and 15\% protons appropriate in compact stellar matter, we get instead
\be
V_{K^-} (\rho=\rho_0)\approx -92\times
(0.85\times \frac 23 +0.15\times \frac 43)
\ \mbox{MeV}\approx -71 \ \mbox{MeV}.
\ee
At $\rho\approx \rho_0$, therefore, the attraction in the stellar matter
of the given composition would be
\be
S_{K^-} +V_{K^-} \approx -171 \ \mbox{MeV}.\label{estimate}
\ee
Normally, a linear extrapolation with density, although commonly
employed in Walecka mean field calculations, is dangerous, because
correlations, Pauli blocking, etc. tend to cut down attraction progressively
with increasing density. However, our effective coupling constant,
starting from chiral Lagrangians, is $1/{f_\pi^\star}^2$, which steadily
increases with density, at least in linear approximation (\ref{GMOR}) -
(\ref{GMOR2}). Given the decrease from many-body effects, and the increase
from the decreasing order parameter $f_\pi^\star$ of the broken chiral
symmetry mode, it is simplest to assume that these effects roughly cancel
each other, and that the mean fields extrapolate linearly in density.
Some empirical evidence for this exists in relativistic heavy-ion
experiments, as we review in the next section.

Given our estimate (\ref{estimate}), we see that
\be
\omega_{K^-} (2\rho_0)\approx 495-342\approx 153\ {\mbox{MeV}}.
\ee
{}From the equation of state for dense stellar matter available in the
literature (e.g., PAL \cite{PAL}), we have the chemical potential for
electrons
\be
\mu_e (2\rho_0)\approx 173 \ {\mbox{MeV}}
\ee
for either $K_0=180$ MeV or 240 MeV.
Since kaon condensation occurs when $\omega_K^\star=\mu_K=\mu_e$,
we expect $\rho_c\sim 2\rho_0$. This agrees well with the $O(Q^3)$
chiral perturbation result $\rho_c\approx 2.3\rho_0$
in \cite{LBMR} when the BR scaling is incorporated. Note, however, that
by using the above linear extrapolation in the mean fields, we have
implicitly assumed a definite parametrization for BR scaling for
$\rho\gsim \rho_0$. We will now argue that there is some justification for
this.
\subsection*{Heavy ion processes}
\indent\indent
To conclude, we review results from analyses of heavy ion reactions.
First of all, in the Au + Au collisions at 1 GeV/N studied at SIS, there
is an enhancement in subthreshold $K^+$ production by a factor of $\sim 3$
due to the attractive scalar potential. Fang et al. \cite{fang} obtain the
enhancement using the scalar interaction (\ref{kaonL}), leaving out
the scaling in $f$ (i.e., setting $f= f_\pi$) and the
factor $F\simeq (1-0.37 \omega_K^2/m_K^2)$. According to the estimate in this
note, these two modifications would very nearly cancel each other.
Maruyama et al. \cite{maruyama} apply the same scalar mean field to the
$\Lambda$-particle as to the nucleon, none to the kaon. However, in
determining the threshold for $K^\pm$ production, which is crucial in
determining the number of subthreshold kaons produced \cite{randrup},
this is equivalent to applying 2/3 of the scalar meson field to the
$\Lambda$, 1/3 to the kaon which we find is a reasonable procedure.
The scalar mean field applied to the kaon is $S_K (\rho_0)\approx -
73$ MeV. This is rather small compared with our recommended value
(\ref{recommend}), $-100$ MeV. However the factor
$F\simeq (1-0.37\omega_K^2/m_K^2)$, with  $\omega_{K^+}\gsim m_K$,
would bring them close together. We thus see that there is some
confirmation of our suggestion of a linear scaling in the mean fields up to
$\rho\lsim 3\rho_0$, the densities relevant for the threshold $K^+$ production.

Another interesting application of the scaling idea is to the
recent dilepton experiments by the CERES collaboration \cite{CERES}.
There the scaling of $f_\pi^\star$ can be related to a medium-dependent
vector meson mass accounting for enhancement of dilepton pairs observed in
200 GeV/N S on Au collisions at the CERN-SPS. The idea of scaled vector masses
has been used successfully in a recent analysis of the CERES process by
Li, Ko and Brown \cite{likobrown}.

The detailed analysis of the SIS and CERES experiments in terms of our
mean-field connection (between the Walecka sector and the kaonic, chiral
sector) will be given in a longer paper \cite{br4}.
\subsection*{Acknowledgments}
\indent\indent
We would like to thank David Kaplan, Chang-Hwan Lee and Tae-Sun Park
for helpful discussions. Part of this work was done while the
authors were participating in the INT95-1 program on ``Chiral
dynamics in hadrons and nuclei" at the Institute for Nuclear Theory,
University of Washington. We would like to acknowledge the hospitality
of the INT and the Department of Energy for partial support.

\end{document}